\documentclass[journal,onecolumn]{IEEEtran}
\IEEEoverridecommandlockouts
\usepackage{cite}
\usepackage{amsmath,amssymb,amsfonts}
\usepackage{algorithmic}
\usepackage{graphicx}
\usepackage{textcomp}
\usepackage{xcolor}
\usepackage{graphicx}
\usepackage{cite}
\usepackage{picinpar}
\usepackage{amsmath}
\usepackage{url}
\usepackage{flushend}
\usepackage[latin1]{inputenc}
\usepackage{colortbl}
\usepackage{soul}
\usepackage{multirow}
\usepackage{makecell}
\usepackage{pifont}
\usepackage{xcolor}
\usepackage{alltt}
\usepackage[hidelinks]{hyperref}
\usepackage{enumerate}
\usepackage{siunitx}
\usepackage{breakurl}
\usepackage{epstopdf}
\usepackage{pbox}
\usepackage{amsfonts}
\usepackage{mathtools}
\usepackage{commath}
\usepackage{booktabs}
\usepackage{caption}
\usepackage{subcaption}
\usepackage{bm}
\usepackage{mathrsfs}

\def\BibTeX{{\rm B\kern-.05em{\sc i\kern-.025em b}\kern-.08em
    T\kern-.1667em\lower.7ex\hbox{E}\kern-.125emX}}

\usepackage{tikz}
\usetikzlibrary{spy}
\usepackage{circuitikz}
\usetikzlibrary{patterns,angles,quotes,intersections,calc}

\DeclareMathOperator*{\argmin}{arg\,min}

\begin{document}

\title{Robust Switching Control of DC-DC Boost Converter for EV Charging Stations \\
\thanks{The authors are with CODIASE Group at LAPLACE (Universit\'e de Toulouse, CNRS, INPT, UPS), Toulouse, France.\\
email: \{ahmad, pitanga, kergus, kader, caux\}@laplace.univ-tlse.fr\\This work is funded by ADEME, the French Agency for Ecological Transition, as part of the I-REVE project.}
}

\author{\IEEEauthorblockN{Saif Ahmad}
\and
\IEEEauthorblockN{Ryan P. C. de Souza}
\and
\IEEEauthorblockN{Pauline Kergus}
\and
\IEEEauthorblockN{Zohra Kader}
\and
\IEEEauthorblockN{St\'ephane Caux}
}
\IEEEoverridecommandlockouts
\maketitle
\IEEEpubidadjcol
\begin{abstract}
In this work, the problem of switching control design for DC-DC boost converter is considered, in the case of operation under uncertain equilibrium condition arising due to perturbations in the input and load parameters. Assuming that these uncertain parameters are generated via a known linear exo-system, a parameter estimator is designed to update the equilibrium point for the switching controller in real-time. In order to mitigate the noise amplification problem associated with the designed parameter estimator, the estimation error injection term is filtered via a set of first-order filters to obtain the desired level of noise suppression in the final set of estimates. To demonstrate the efficiency of the developed scheme, a realistic application scenario of a DC charging station for electric vehicles is considered, with photovoltaic array as the source and a battery connected at the load side.

\end{abstract}


\section{Introduction}

Microgrids are small-scale grids that integrate distributed loads and sources, able to operate with or without the contribution of the utility grid, providing significant benefits to the connected users as well as to the utility grid \cite{kumar2017DCmicrogrid}. Comparison between AC and DC microgrids is available in \cite{justo2013ac}, recalling that, with the projected increase in the number of DC powered components and of distributed sources that generate DC power, DC microgrids are a good candidate for future energy systems. Indeed, DC microgrids offer better compatibility with distributed renewable energy resources, higher reliability and efficiency \cite{kumar2017DCmicrogrid}. In particular, they are increasingly popular for Electrical Vehicle (EV) charging \cite{kaur2017state}. Indeed, due to their capacity to absorb intermittent generation from renewable energy sources, DC microgrids allow to mitigate the burden and negative impacts of EV charging on the electrical grid.

In order to meet the expectation regarding the performance of such DC microgrids, efficient control and energy management are required. As recalled in \cite{dahale2017overview}, there are three hierarchical control levels for DC microgrids: primary, secondary and tertiary control. At the highest level, \textit{tertiary control}, also known as Energy Management System (EMS), schedules the power dispatch between source and storage. The source load is then shared among the different sources through \textit{secondary control}. This is done by sending references to be tracked by \textit{primary control}, consisting of the local controllers, enabling proper load sharing for parallel connected converters. At each control level, the performance depends on the accuracy and performance of the following control level, which makes DC-DC converter control of crucial importance. 

In this context, a number of linear control strategies have been proposed in literature based on averaged small signal models of DC-DC converters  \cite{kobaku2017experimental,kobaku2020quantitative}, which ensures desired nominal performance and system stability but only in the neighbourhood of the equilibrium point. However, power converters are generally characterised by highly nonlinear dynamics and continuous parametric variations subject to source and load conditions, which in turn shifts the equilibrium point. This highlights the limitations of closed-loop performance achievable via linear feedback control and the need for more advanced control strategies in this area that account for the switching and parameter-varying nature of converters. Advanced control methods for DC-DC converters in DC microgrids are reviewed in \cite{xu2020review} and classified in the following categories: model predictive control, backstepping, sliding mode control (SMC), passivity-based control, observer/estimation based technique and intelligent control. 
More recently, principles of switched systems theory has been successfully employed as an alternative to the averaging approach for control of power converters in \cite{de2022switching}. One of the main advantages of using switched systems theory is that linearization around the equilibrium is not required, and global stabilization can be ensured. Furthermore, a modulation stage, such as Pulse-Width Modulation (PWM), is not required either.

However, the switching controller in \cite{de2022switching} assumes that the input and load parameters remain unchanged for the power converters, which in turn results in a  constant equilibrium point. On the other hand, in most practical scenarios such as DC-microgrids and EV charging stations, power electronic converters are subject to time-varying source and load perturbations \cite{bayati2020sinusoidal}, resulting in continuous variation of the operating point. This problem has recently been addressed for DC-DC power converters by using augmented state observer \cite{beneux2019adaptive} and state augmentation via tracking error integral \cite{ndoye2022switching}. In the present work, we address the problem of uncertain equilibrium in the context of switching control designed for a DC-DC boost converter (DBC) by estimating the uncertain parameters. The developed solution results in the following advantages: \textbf{(a).} Unlike \cite{beneux2019adaptive}, the designed estimator approximates only the required unknown parameters and not the measurable states. Moreover, noise amplification problem commonly associated with observers/estimators (such as those used in \cite{beneux2019adaptive}) is addressed by the estimation error filtering approach introduced in \cite{astolfi2021use} which results in better estimation quality and allows the freedom to introduce  desired level of filtering in the obtained estimates. Also, unlike \cite{ndoye2022switching}, the parameter estimation approach returns physical quantities (input voltage and load current) that are important from an operational perspective (e.g. load sharing via droop control) in an interconnected network of power converters. It is also worth mentioning that the approach used in \cite{ndoye2022switching} is designed for a buck converter which is much simpler to handle while the present work considers a boost converter which has nonlinear dynamics. \textbf{(b).}  The resulting estimation error dynamics are devoid of switched terms which in turn facilitates the tuning of estimator, and gives simpler \textit{linear matrix inequality} (LMI) conditions for stability of the designed estimator.  Furthermore, the robust switched control algorithm is implemented via a hysteresis block which provides a simple and intuitive solution to the infinite switching frequency problem \cite{de2022switching} and facilitates practical implementation. In contrast to the previous works which consider constant resistances at the load end and step changes in input voltages, the present work considers a PV array connected at the input side and a Li-ion battery charging at the load side in order to simulate a more practical operating condition.



Remaining sections in this paper are organized as follows. Section \ref{sec:problem_formulation} first introduces the switching model for boost converters and highlights the importance of considering variations of operating points in control design of DC-DC converters. Section \ref{sec:control_design} then exposes the proposed control and estimation algorithm along with the noise filtering and hysteresis approach to facilitate practical implementation. The efficiency of the proposed approach is then illustrated in Section \ref{sec:numerical_simulation} through numerical simulations. The paper concludes in Section \ref{sec:conclusion} with a summary of the results and future research directions.


\subsubsection*{Notations} $\bm{I_n}$ is an identity matrix of dimension $n\times n$, $\textbf{0}$ is a zero vector or matrix of appropriate dimension, $\textrm{diag}(a_1,\dots,a_n)$ denotes a diagonal matrix with $a_1,\dots,a_n$ as diagonal elements.





\section{Problem Formulation}
\label{sec:problem_formulation}

Fig. \ref{fig:boost} shows the ideal circuit of a DC-DC boost converter (DBC) where $v_{in}$ is the source voltage, $R_o$ is a load resistance, $i_{Load}$ denotes the current being supplied to another load connected in parallel, $L$ and $C$ are the inductance and capacitance respectively while $i_L$ and $v_o$ are the system states representing inductor current and output voltage respectively. We define $\sigma\in\{0,1\}$ as a switching variable which  corresponds to  the state of the switched device $S$ such that $\sigma=1$ when switch $S$ is on and $\sigma=0$ when the switch is in off state. The dynamics of a DBC can now be defined with the following expression:
\begin{figure}
    \centering
    \begin{center}
        \scalebox{0.9}{
            \begin{circuitikz}[american voltages] 
                \draw
                (0,0) to [american voltage source, v<=$v_{in}$,invert] (0,3)
                to [L,i=$ i_L $,l=$ L $] (2,3)
                to [empty diode] (4,3)
                to [short](6,3)
                to [R,l=$R_o$] (6,0)
                (0,0) to [short] (8,0)
                (6,3) to [short](8,3)
                
                (2,3) to [closing switch,l_=$ S $] (2,0)
                (4,3) to [C,l=$ C $,v=$ v_o $] (4,0)
                (8,3) to [generic,i=$i_{Load}$] (8,0);
        \end{circuitikz}}
    \end{center}
    \caption{Circuit of a DC-DC boost converter.}
    \label{fig:boost}
\end{figure}
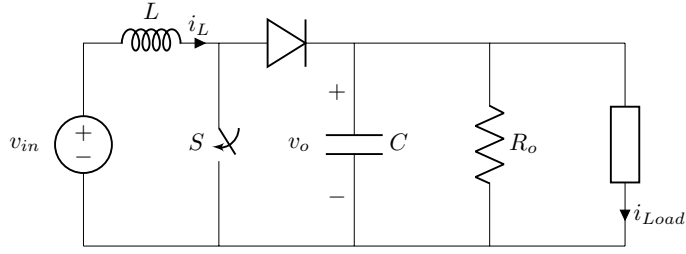
\begin{equation}\label{eq:system}
    \begin{split}
         \dot{\bm{x}}(t)=\bm{A_{\sigma}x}(t)+\bm{G}\bm{p}(t)
    \end{split}
\end{equation}
    where $\bm{x}(t):=[i_L(t),v_o(t)]^T, \ \bm{p}(t)=[p_1(t),p_2(t)]^T:=[v_{in}(t),i_{Load}(t)]^T,$
  \begin{equation}
      \begin{split}
          \bm{G}=&\begin{bmatrix}\frac{1}{L}&0\\ 0&-\frac{1}{C}\end{bmatrix},\bm{A_{\sigma}}=\begin{bmatrix}0&-\frac{(1-\sigma)}{L}\\ \frac{(1-\sigma)}{C}& -\frac{1}{R_oC}\end{bmatrix}.
      \end{split}
  \end{equation}


It can be shown that the possible equilibria of the instantaneous model \eqref{eq:system} are the same as those of the averaged model \cite{beneux2019adaptive}, whose dynamics are given by:
\begin{equation}\label{eq:averagedModel}
    \bm{\dot{x}}(t) = \bm{A}(\bar{\sigma})\bm{x}(t) + \bm{Gp}(t),
\end{equation}
where $\bm{A}(\bar{\sigma}) = \bar{\sigma} \bm{A_1} + (1-\bar{\sigma})\bm{A_0}$.

The control objective in this paper is to stabilize the output voltage at a desired equilibrium value $v_o^\star$. A point $\bm{x^*} = [i_L^\star\;\;v_o^\star]^T$ is an equilibrium of \eqref{eq:averagedModel} if there exists a $\sigma^*\in[0,1]$ such that $\bm{A}(\sigma^*)\bm{x^*} + \bm{Gp} = \textbf{0}$ for a given $\bm{p}(t)$. Here, the value of $\bar{\sigma}^*$ is the steady-state duty cycle. Assuming $\bar{\sigma}^* > 0$, $\bm{x}^*$ can be expressed as $\bm{x}^* = -\bm{A}(\bar{\sigma}^*)^{-1}\bm{Gp}$. 
\section{Robust Switching Control Design}
\label{sec:control_design}
It can easily be checked that \eqref{eq:system} is an example of a class of systems known as \textit{switched affine systems} (SAS). This means that, for each $\sigma \in \{0,1\} $, the dynamics are expressed as an affine function of the state $\bm{x}$. Several works in the literature deal with control design for these systems. In the next subsection, one of these strategies ensuring global stabilization of the equilibrium point is presented.
\subsection{Switching Controller}\label{subsec:switchingLaw}
Assuming that $\bm{x^*}$ is perfectly known, the SAS defined by \eqref{eq:system} can be asymptotically stabilized to $\bm{x^*}$ if for a given $\alpha$ such that $0\leq\alpha\leq\bar{\alpha}$ (where $\bar{\alpha}>0$ is dependent on the open-loop system matrix $A(\sigma^*)$), there exists $\bm{P}=\bm{P}^T\succ0$ such that
\begin{equation}\label{eq:LMIforP}
    \bm{A}^T(\sigma^*)\bm{P}+\bm{PA}(\sigma^*)+2\alpha\bm{P}\leq 0.
\end{equation}

The preceding LMI imposes an implicit decay rate $\alpha$ on the tracking error via appropriate selection of $\bm{P}$. The feasibility of \eqref{eq:LMIforP} is a convex problem, and efficient solvers are widely available. 

It has been shown in \cite{Bolzern:2004} that, once $\bm{P}$ is found, the following switching controller stabilizes system \eqref{eq:system}:
\begin{equation}\label{eq:switchingLaw1}
    \sigma(\bm{x}(t)) \in \argmin_{m\in\{0,1\}} (\bm{x}(t)-\bm{x^*})^T \bm{P} (\bm{A_m x}(t) + \bm{G} \bm{p}(t)).
\end{equation}
Since the term $ \bm{G p}(t) $ does not depend on the mode $\sigma$ in DBC, \eqref{eq:switchingLaw1} can be rewritten as:
\begin{equation}\label{eq:switchingLaw}
    \sigma(\bm{x}(t)) \in \argmin_{m\in\{0,1\}} (\bm{x}(t)-\bm{x^*})^T \bm{P A_m x}(t).
\end{equation}
As shown in \cite{Bolzern:2004}, control law \eqref{eq:switchingLaw} globally stabilizes system \eqref{eq:system}. It is worth noting that, unlike more traditional control methods, the switching law presented here does not require linearization around the equilibrium point. Even so, it ensures global stabilization. Another practical advantage is that the controller directly outputs the mode to be selected, thereby bypassing the need for a modulation stage between the controller and the switching device.

However, it can easily be seen that the controller \eqref{eq:switchingLaw} requires the knowledge of $\bm{x}^*$, which, as previously established, is given by $\bm{x}^* = -A(\sigma^*)^{-1}G\bm{p}(t)$. It can be checked that $\bm{x}^*$ depends in fact on both $p_1(t)$ and $p_2(t)$ for the case of DBC:
\begin{equation}\label{eq:xstar}
    x^\star = \frac{1}{1-\sigma^*}\left[\begin{array}{c} p_2 \\ p_1 \end{array}\right].
\end{equation}
Note from \eqref{eq:xstar} that $\sigma^*$ can be determined as:
\begin{equation}\label{eq:sigma*}
    \sigma^* = 1 - \frac{p_1}{v_o^\star}.
\end{equation}

Therefore, in order to implement \eqref{eq:switchingLaw}, the parameter vector $\bm{p}(t)$ must be known. In addition, solving LMI \eqref{eq:LMIforP} requires the knowledge of $\sigma^*$, which depends on $p_1(t)$. However, $\bm{p}(t)$ itself is uncertain and depends upon the source as well as load conditions which in turn necessitates either the measurement of these additional signals (which incurs additional cost) or estimation in order to ensure accurate tracking of the desired uncertain and possibly time-varying equilibrium point $\bm{x^*}(t)$, where the variation with respect to time is sufficiently slow. A straightforward implementation of \eqref{eq:switchingLaw} requires the equilibrium point to be updated at each time-step using the measurement of $\bm{p}(t)$. If the measurements are unavailable, it becomes necessary to estimate $\bm{p}(t)$, which is the subject of the next subsection. 
\subsubsection*{Remark 1} SAS defined  in \eqref{eq:system} and the corresponding switching controller in \eqref{eq:switchingLaw1} is not specific to a DBC and can be used to model and control a wider class of DC-DC converters including buck, buck-boost, flyback converters among others \cite{beneux2019adaptive,ndoye2022switching}. Moreover, the problem of uncertain equilibrium is common to all power converters due to the inherent intermittency of the renewables and uncertain load conditions (time-varying $\bm{p}(t)$) in most applications. Hence, the presented solutions, results and ensuing discussions are more general and can easily be extended for other types of converters.
\subsection{Parameter Estimator}
Since we assume that now the parameters $\bm{p}$ may change with time, then, as mentioned before, the equilibrium point $\bm{x^*}$ changes accordingly. Let us assume that $p_1$ is bounded in the following way: $v_{in}^- \leq p_1 \leq v_{in}^+ < v_o^\star$, where $v_{in}^-$ and $v_{in}^+$ define the range in which the input voltage may evolve and they depend on its uncertainty. Note that if $v_{in}^+ > v_o^\star$, then the equilibrium point may be unreachable, since the output voltage of a DBC cannot be lower than the input voltage. As $v_{in}$ evolves over time, the steady-state duty cycle $\sigma^*$ varies according to \eqref{eq:sigma*}.

In order for the control law \eqref{eq:switchingLaw} to stabilize the system in all the range of possible values for $\sigma^*$, one must find a single matrix $\bm{P}$ such that \eqref{eq:LMIforP} holds for all $\sigma^* \in [\sigma_{min}^*,\sigma_{max}^*]$, where $\sigma_{min}^* = 1-v_{in}^+/v_o^\star$ and $\sigma_{max}^* = 1-v_{in}^-/v_o^\star$. Thanks to the convexity of \eqref{eq:LMIforP}, this is true if the following two LMIs hold:
\begin{equation}\label{eq:LMIforPmod}
    \bm{A}^T(\mu)\bm{P}+\bm{PA}(\mu)+2\alpha\bm{P}\leq 0,\quad \mu \in\{\sigma_{min}^*,\sigma_{max}^*\}.
\end{equation}

Now, in order to obtain the estimate of uncertain parameters, it is assumed that the parameter vector $\bm{p}(t)$ is generated by a known linear exo-system of the form
\begin{equation}
    \begin{split}
        \bm{\dot{\zeta}_p}(t)&=\bm{A_p\zeta_p}(t)\\
        \bm{p}(t)&=\bm{C_p\zeta_p}(t),
    \end{split}
    \label{eq:dist_dynamics}
\end{equation}
where  $\bm{\zeta_p}\in\mathbb{R}^{m}$ is the state-vector defining the exo-system while $\bm{A_p}\in \mathbb{R}^{m\times m}$ and $\bm{C_p}\in \mathbb{R}^{2\times m} $ are the associated matrices. The preceding assumption allows us to consider a wider class of parameter behaviors such as linearly varying with time, sinusoidal etc. or a combination, that arise in practical systems. Furthermore, the common assumption of $\bm{\dot{p}}=\textbf{0}$ considered in \cite{beneux2019adaptive,ndoye2022switching}, is a special case of \eqref{eq:dist_dynamics} obtained with $\bm{A_p}=\textbf{0}$ and $\bm{C_p}=\bm{I_2}$, which is also used in the current work. The following switched parameter estimator can now be constructed upon \eqref{eq:system} and \eqref{eq:dist_dynamics} following the general approach introduced in \cite{chen2004disturbance}:
\begin{equation}
\begin{split}
    \dot{\hat{\bm{z}}}_{\bm{p}}(t)=&(\bm{A_p}-\bm{\kappa GC_p})\bm{\hat{z}_p}(t)+(\bm{A_p-\kappa GC_p})\bm{\kappa}\bm{x}(t)-\bm{\kappa}\bm{A_{\sigma}}\bm{x}(t)\\
   \hat{\bm{\zeta}}_{\bm{p}}(t) =&\hat{\bm{z}}_{\bm{p}}(t)+\bm{\kappa x}(t)\\
   \hat{\bm{p}}(t)=&\bm{C_p\hat{\zeta}_p}(t)
\end{split}
\label{eq:pe}
\end{equation}
where $\bm{\hat{z}_p}\in\mathbb{R}^{m}$ is an intermediate variable (introduced to avoid the implementation of derivative $\bm{\dot{x}}(t)$), $ \bm{\hat{\zeta}_p}$ and $\bm{\hat{p}}$ are estimates of $\bm{\zeta_p}$ and $\bm{p}$ respectively, while $\bm{\kappa}\in \mathbb{R}^{m\times2}$ is the observer gain vector that needs to be selected.

From \eqref{eq:dist_dynamics} and \eqref{eq:pe}, the error dynamics for the designed estimator is obtained as 
\begin{equation}
    \bm{\dot{e}}(t)=(\bm{A_p}-\bm{\kappa G C_p})e(t)
    \label{eq:estimation_error}
\end{equation}
where $\bm{e}(t):=\bm{\hat{\zeta}_p}-\bm{\zeta_p}$ (and consequently $\bm{e_p}(t):=\bm{p}(t)-\bm{\hat{p}}(t)=\bm{C_pe}(t)$) converges asymptotically if $\bm{\kappa}$ is selected such that $\bm{A_p}-\bm{\kappa G C_p}$ is Hurwitz, which is equivalent to saying that there exists a $\bm{\Bar{P}}=\bm{\Bar{P}}^T\succ\textbf{0}$, such that
\begin{equation}
    (\bm{A_p}-\bm{\kappa G C_p})^T\bm{\Bar{P}}+\bm{\Bar{P}}(\bm{A_p}-\bm{\kappa G C_p})\leq \textbf{0}.
    \label{eq:LMI_estimator_stability}
\end{equation}Condition \eqref{eq:LMI_estimator_stability} is obtained readily upon selection of a Lyapunov function of the form $V=\bm{e}^T\bm{\bar{P}e}$ and taking its derivative along the trajectories of estimation error dynamics \eqref{eq:estimation_error}, following the approach given in \cite{chen2004disturbance}. In this case, it can be shown that the estimation error dynamics is \textit{globally asymptotically stable}.

\subsubsection*{Remark 2} For the sake of simplicity, it is assumed in this paper that $R_o$ is a known load resistance. However, uncertainties in this parameter can be addressed by adapting the LMIs in \eqref{eq:LMIforPmod} to be solved in a similar way as in \cite{beneux2019adaptive}, for instance. 
\subsubsection*{Remark 3} In a practical scenario, it is not possible to completely characterize the variations in uncertain parameters with a known linear exo-system and hence, it is more appropriate to consider the exo-system dynamics as 
\begin{equation}
    \begin{split}
        \bm{\dot{\zeta}_p}(t)&=\bm{A_p\zeta_p}(t)+\mathcal{H}\\
        \bm{p}(t)&=\bm{C_p\zeta_p}(t),
    \end{split}
    \label{eq:practical_dist_dynamics}
\end{equation}
where $\mathcal{H}\in\mathbb{R}^m$ represents the residual of the exo-system i.e. the model mismatch. The convergence results obtained previously can be extended for the practical scenario considering \eqref{eq:practical_dist_dynamics} to show that that the estimation error dynamics is \textit{input to state stable} (ISS) \cite{astolfi2021use} with respect to $\mathcal{H}$, provided $\|\mathcal{H}\|\leq\mu$ where $\mu$ is a positive constant, which is always the case in practical applications due to physical limitations of real systems and implementation of protection mechanisms. Under a high-gain formulation \cite{astolfi2021use}, it is also possible to show that the region in which the estimation error eventually converges under the effect of $\mathcal{H}$ can be made made arbitrary small and the rate of convergence can be made arbitrarily fast by increasing the high-gain parameter \cite{ahmad2021active}. We apply this approach for tuning the estimator parameter $\bm{\kappa}$ in the current work such that all the eigenvalues of $\bm{A_p}-\bm{\kappa G C_p}$ are placed at $``-\lambda"$ which denotes the high-gain parameter in this case. This in turn reduces the number of parameters to be selected and simplifies the tuning process to facilitate practical implementation.
\subsubsection*{Remark 4} As shown in \cite{ahmad2021active}, it is possible to obtain a more accurate estimate of time-varying parameters by approximating their dynamics using a $k^{th}$ order time-polynomial form. However, the noise amplification associated with high-gain construction  is aggravated in this case and additional filtering becomes necessary \cite{ahmad2021active,astolfi2021use}. 
\subsection{Noise Filtering}
In real systems, state measurements are always corrupted by high-frequency noise component $(\nu)$ which gets added during the measurement process. Therefore, the designed estimator and controller receive corrupted state measurements $(\bm{x_m}:=\bm{x}+[1,1]^T\nu)$ instead of actual states. On close inspection of \eqref{eq:pe}, it becomes evident that the estimate $\bm{\hat{p}}(t)$ is directly affected by noise in state measurements \textit{i.e.} there is no filtering in the designed estimator. Furthermore, a high value value of $\bm{\kappa}$ (obtained by selecting a high value of $\lambda$), which is necessary for tracking fast variations in the source and load parameters, directly amplifies the noise component in these measurements thereby corrupting the obtained estimates and limiting the practical application of the developed scheme particularly in noisy environment. In order to address the noise amplification problem, we modify the estimator dynamics following the approach introduced in \cite{astolfi2021use} such that the estimation error injection term is filtered via a set of first-order filters. The resulting estimator dynamics are obtained in the following expression:
\begin{equation}
\begin{split}
 \bm{\dot{\hat{\zeta}}_p}(t)&=\bm{A_p\hat{\zeta}_p}(t)+\bm{\kappa} \bm{z_r}(t) \\
\bm{\dot{\eta}}(t)&=\bm{\theta}\big[-(\bm{A_{\sigma}}+\bm{\theta})\bm{x}(t)-\bm{GC_p\zeta_p}(t)-\bm{\eta}(t)\big]\\
\bm{\dot{z}_i}(t)&=\bm{\theta}\big[\bm{z_{i-1}}(t)-\bm{z_i}(t)\big], \ i=\{2,\dots,r\}\\
\bm{z_1}(t)&=\bm{\eta}(t)+\bm{\theta x}(t)\\
\bm{\hat{p}}(t)&=\bm{C_p\hat{\zeta}_p}(t)\\
\end{split}
\label{eq:estimator_noise}
\end{equation}
where $\bm{\theta}=\textrm{diag}(\lambda_{\theta},\lambda_{\theta})$ is the filter gain vector parameterized in terms of $\lambda_{\theta}$. In order to ensure that the accuracy of the designed estimator remains unaffected, the filter parameter is selected such that $\lambda_{\theta}=\gamma\lambda$ where $\gamma>1$ is a tuning parameter for the filters. It is to be noted that $r$ is a design choice which decides the degree of filtering in the obtained estimates by introducing a relative degree of $r$ between the estimate $\bm{\hat{p}}(t)$ and the noise $(\nu)$ entering via state measurements.
\subsection{Hysteresis-based switching} \label{Section:Hysteresis}
One important drawback of the switching law \eqref{eq:switchingLaw} is that it may lead to a sliding mode \cite{Bolzern:2004}, which in practical terms means that the switching frequency rises to unacceptably high levels. Note that, according to \eqref{eq:switchingLaw}, switching only occurs when both modes minimize the objective function in \eqref{eq:switchingLaw}. This can only occur on the surface $s(x) = 0$, where $s(x):=(x - x^\star)^T P D x$ and $D:= A_1 - A_0$.

Here, an hysteresis-based strategy is employed to bound the switching frequency at a finite value. In this manner, switching is not allowed while $|s(x)| < h$, with $h > 0$ a parameter to be determined.

Assuming that the switching frequency is high compared to the system bandwidth (which often holds in practice since the same assumption is made when pulse width modulation (PWM) is used), it can be shown \cite{de2022switching} that the steady-state switching frequency $f_s(p)$ is determined as:
\begin{equation}\label{eq:switchingFrequency}
    f_s(p) = \frac{1}{2h(p)}\frac{|b_0^T P Dx^\star {x^\star}^TD^T P b_1|}{\left(|b_0^T P Dx^\star| + |b_1^T P Dx^\star|\right)},
\end{equation}
where $b_\sigma := A_\sigma x^\star + Gp $, $\sigma = 0,1$. Since the steady-state switching frequency also depends on the parameters $ p $, then their estimation can also be used to update the hysteresis width $ h(p) $ in order to obtain the user-specified switching frequency $ f_s(p) $ using \eqref{eq:switchingFrequency}.

\section{Numerical Simulation}
\label{sec:numerical_simulation}
Simulation study is carried out in  Simulink environment of MATLAB with the simulation parameters listed in Table \ref{table:sim_para}. The switching frequency of the implemented control scheme was limited to $200$kHz via the hysteresis approach defined in Section \ref{Section:Hysteresis}. A high-pass filtered white noise is added to the state variables  going to the estimator designed in \eqref{eq:estimator_noise}, in order to replicate the effect of measurement noise $(\nu)$. Nominal circuit, controller and estimator parameters used for the simulation study are also presented in Table \ref{table:sim_para}. Estimator parameter $\lambda$ is tuned to be faster than the convergence rate $(\alpha)$ for the switching controller while the filter parameter is selected as $\gamma=2.5$ to ensure appropriate filtering of the noise component without affecting the estimator performance. Sufficient filtering of the estimates is obtained with a unity relative degree between measurement noise and the estimates by selecting $r=1$ in \eqref{eq:estimator_noise}. For the simulation study, three different test scenarios are considered which are discussed next. 
\begin{table}[]
    \centering
    \begin{tabular}{|p{30mm}|p{130mm}|}
    \hline
        Circuit& $L=1mH, C=50\mu F, R_o=10\Omega, i_{Load}=0A,v_{in}=350V,v_{in}^+=400V,v_{in}^-=300V,v_o^*=450V$  \\
        \hline
        Controller \& Estimator & $\alpha=40,\bm{A_p}=\textbf{0}$, $\bm{C_p}=\bm{I_2}$, $\lambda=4000,\gamma=2.5,r=1$
         \\
         \hline
         Simulation & $T_s=1/100f_s$, $f_s=200$kHz, Noise $(\nu)$= High-pass filtered (100kHz cut-off) white noise (power 1e-10) \\
         \hline
    \end{tabular}
    \caption{Parameters used for the simulation study.}
    \label{table:sim_para}
\end{table}
\begin{figure}[]
     \centering
     \begin{subfigure}[b]{0.3\textwidth}
              \centering
         \includegraphics[width=\textwidth]{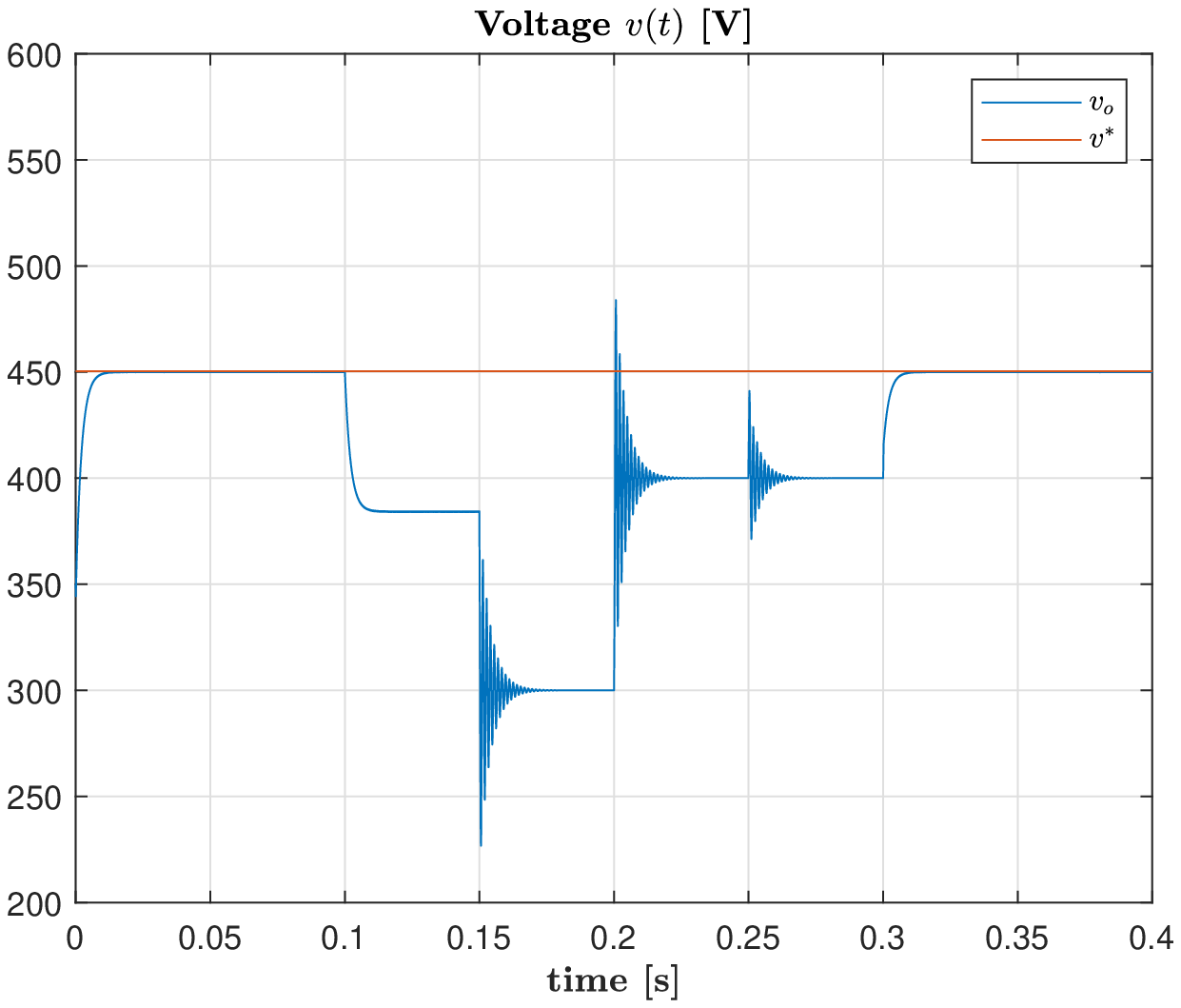}
         \caption{ Output voltage}
         \label{fig:S1_voltage}
     \end{subfigure}\hspace{15mm}
       \begin{subfigure}[b]{0.3\textwidth}

     \centering
         \includegraphics[width=\textwidth]{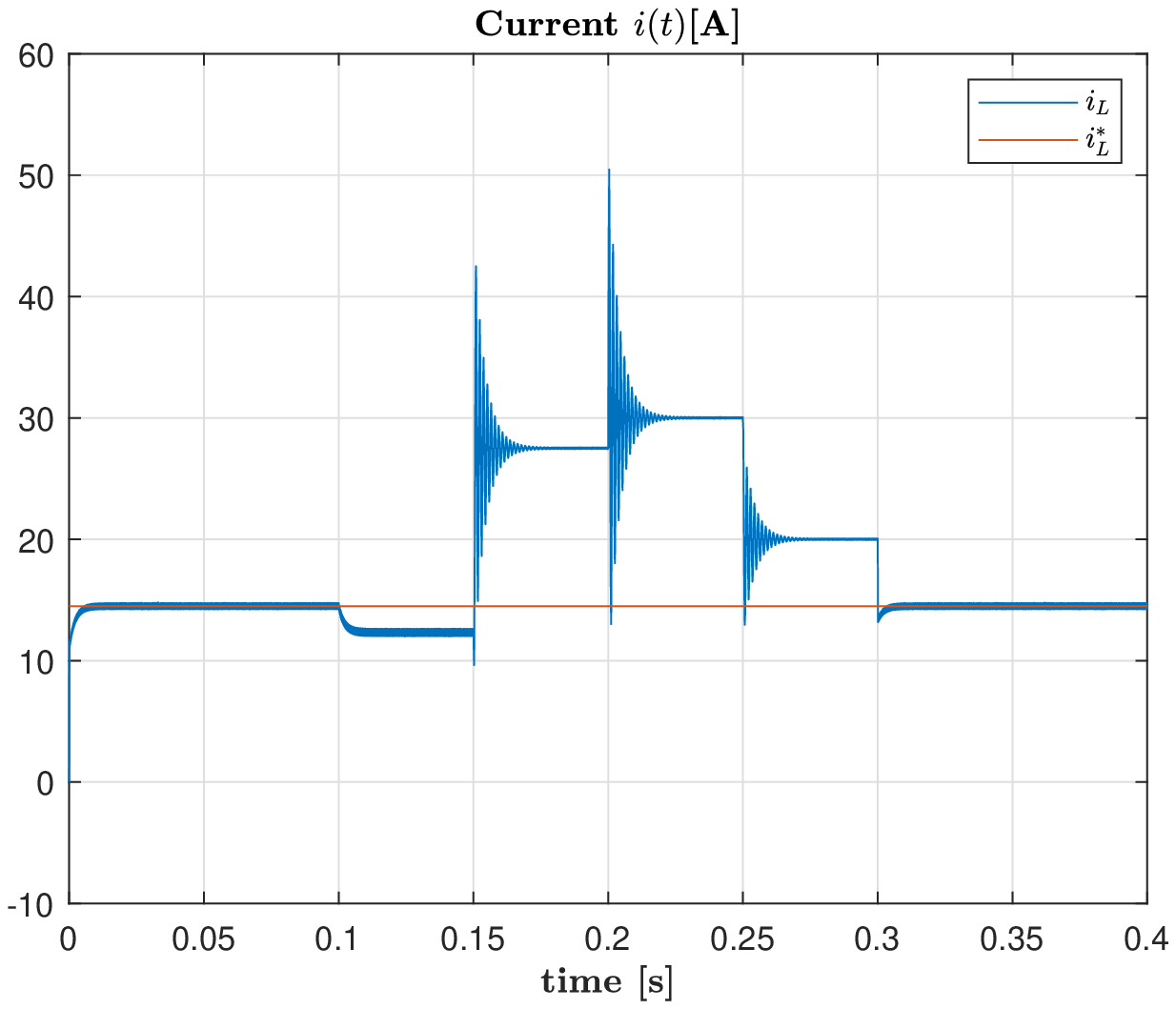}
         \caption{ Inductor current}
         \label{fig:S1_current}
     \end{subfigure}
        \caption{ Simulation plots for \textbf{\textit{S1}}.}
        \label{fig:S1}
\end{figure}

\begin{figure}[]
     \centering
     \begin{subfigure}[b]{0.3\textwidth}
              \centering
         \includegraphics[width=\textwidth]{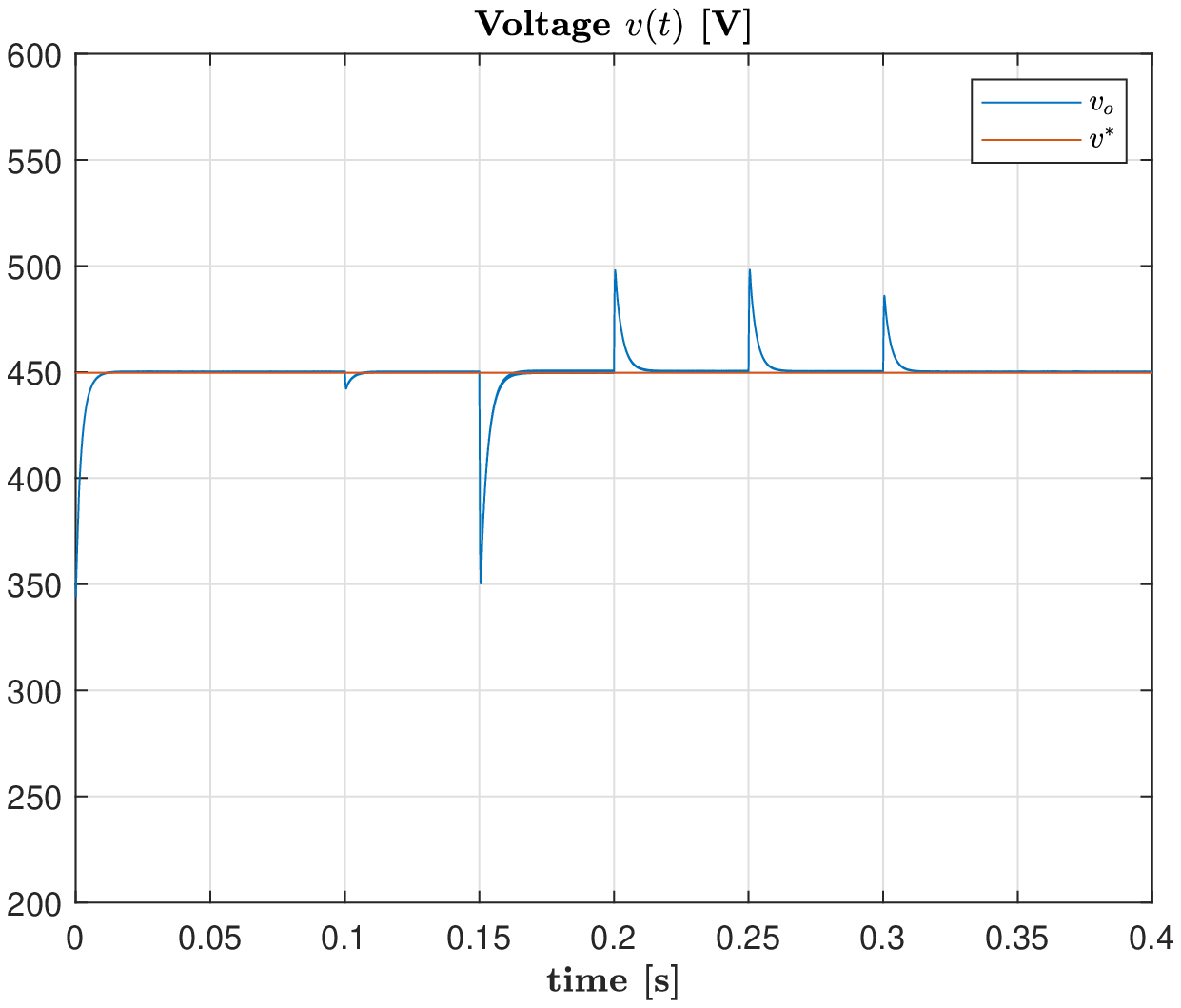}
         \caption{Output Voltage}
         \label{fig:S2_output_voltage}
     \end{subfigure}\hspace{15mm}
       \begin{subfigure}[b]{0.3\textwidth}
     \centering
         \includegraphics[width=\textwidth]{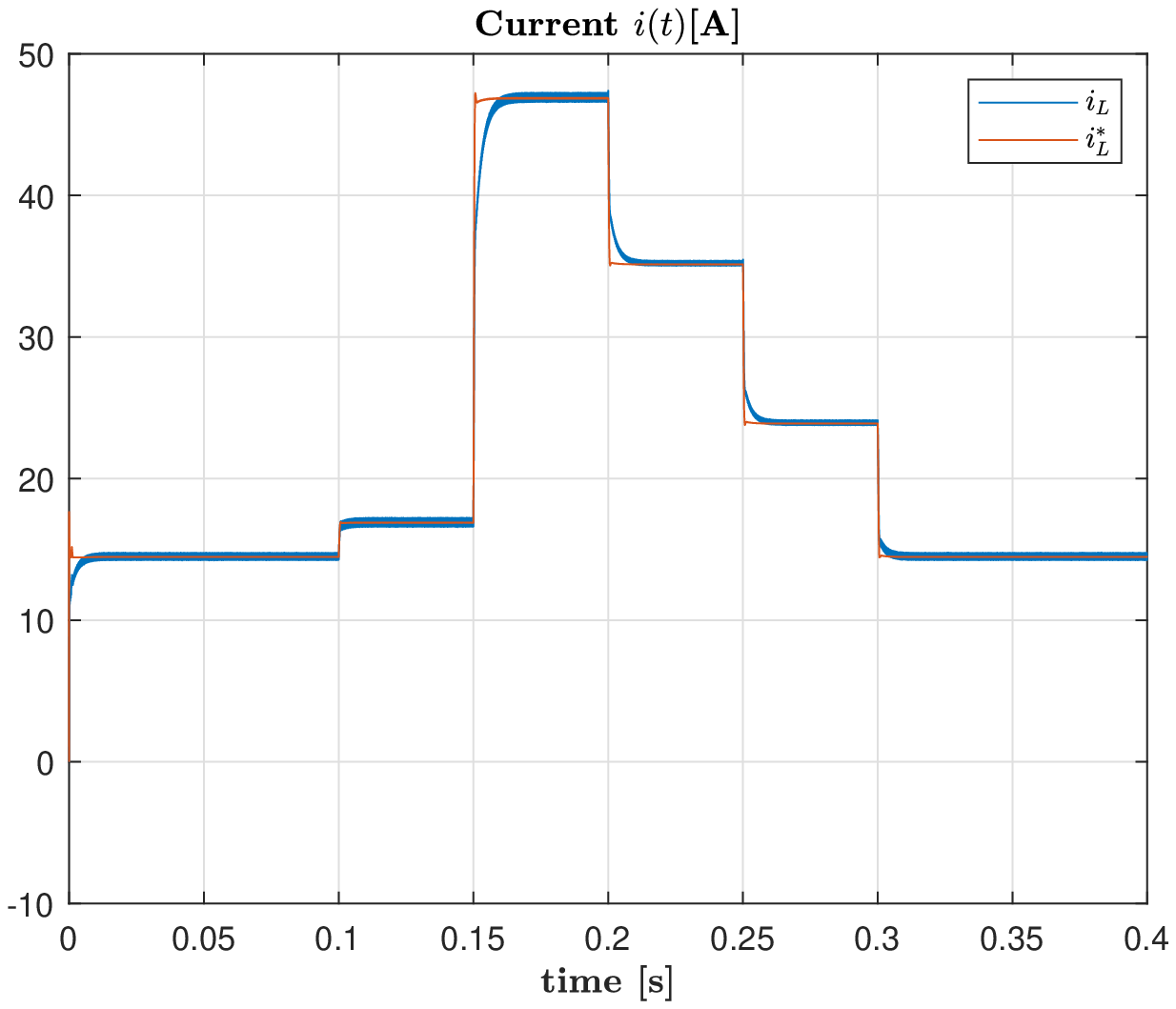}
         \caption{ Inductor current}
         \label{fig:S2_inductor_current}
     \end{subfigure}
          \\
          \begin{subfigure}[b]{0.3\textwidth}
         \centering
         \includegraphics[width=\textwidth]{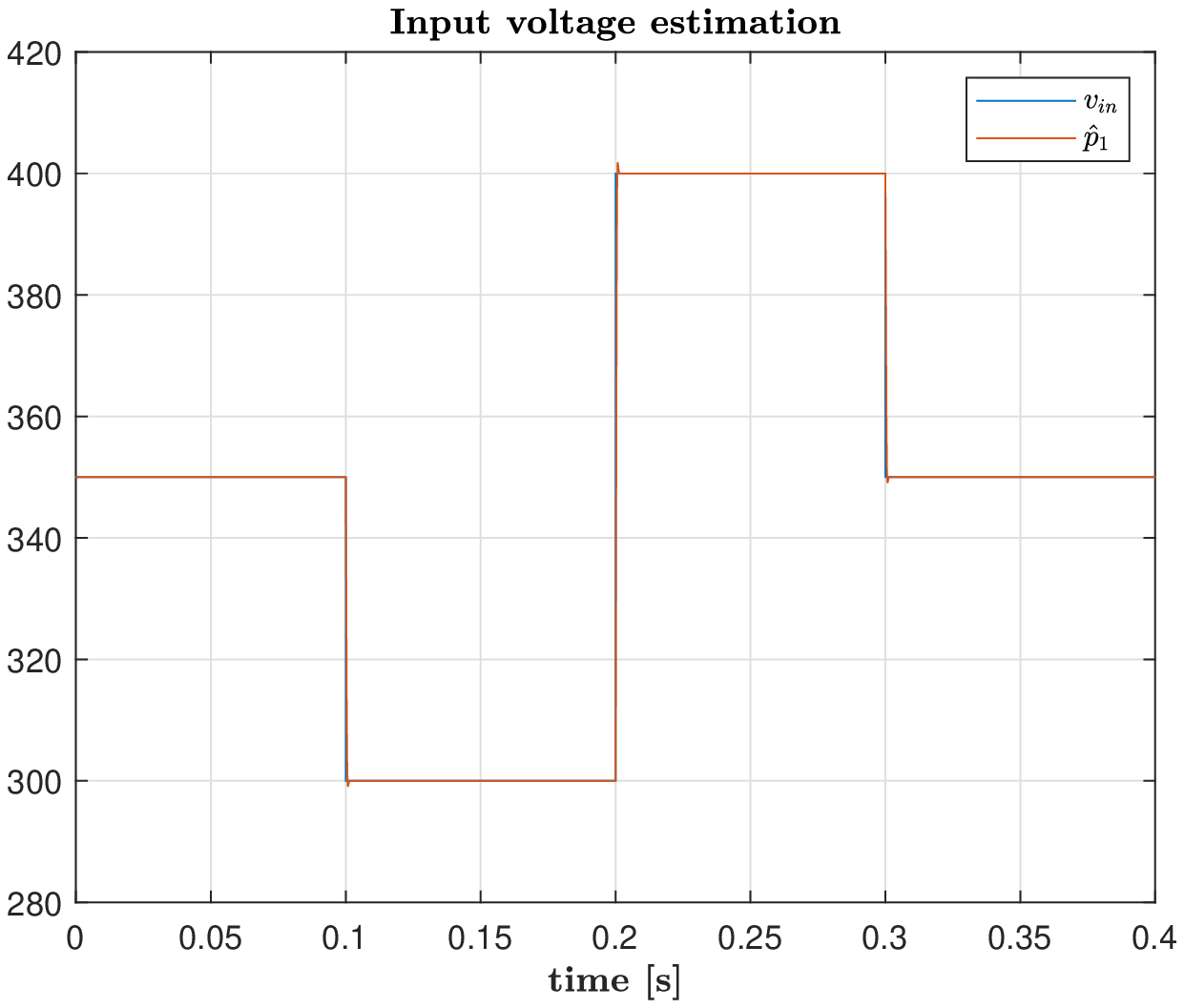}
         \caption{$v_{in}$ and $\hat{p}_1$}
         \label{fig:S2_p1_est}
     \end{subfigure}\hspace{15mm}
               \begin{subfigure}[b]{0.3\textwidth}
         \centering
         \includegraphics[width=\textwidth]{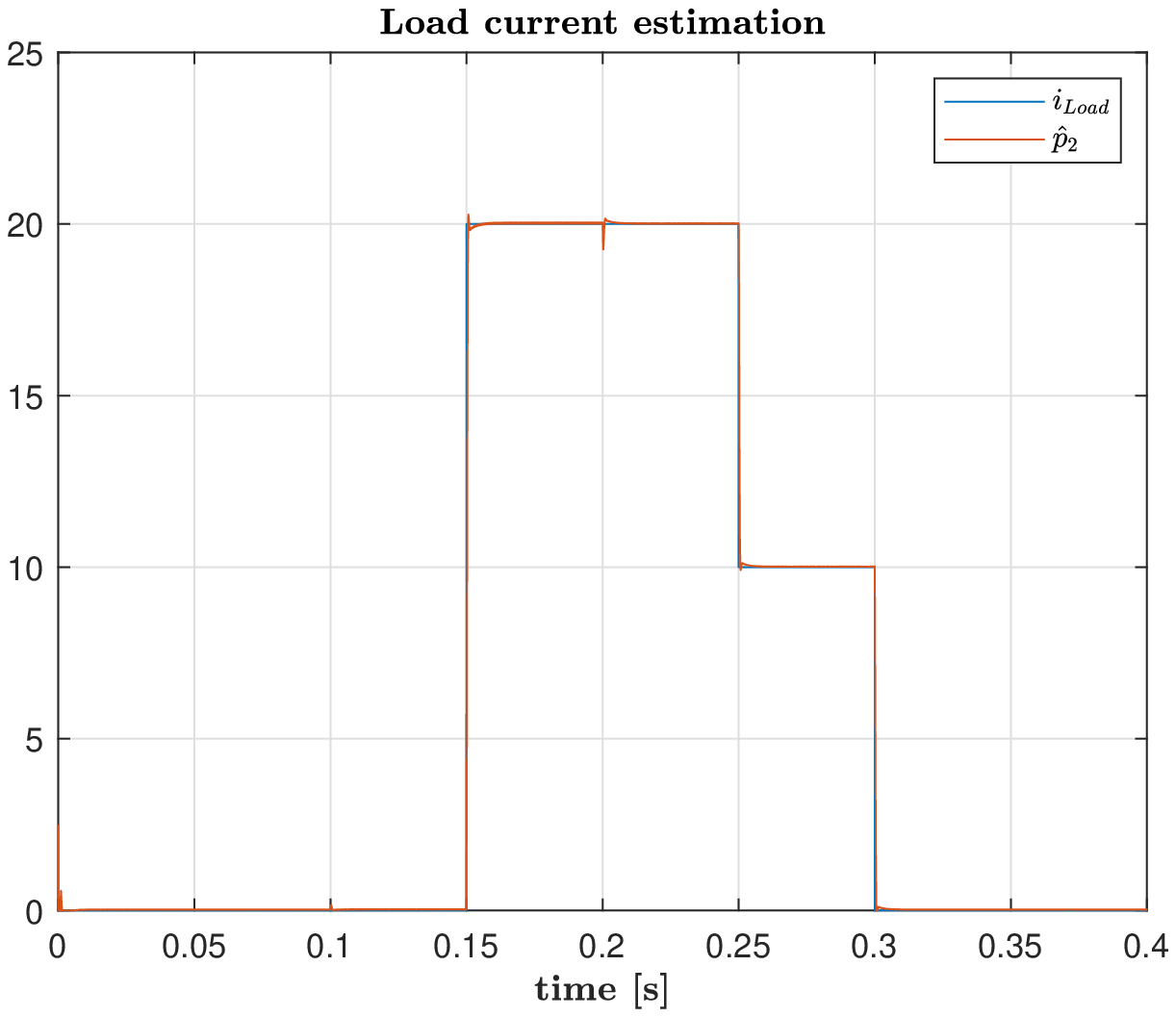}
         \caption{$i_{Load}$ and $\hat{p}_2$}
         \label{fig:S2_p2_est}
     \end{subfigure}
        \caption{Simulation plots for \textbf{\textit{S2}}.}
        \label{fig:S2}
\end{figure}

\begin{figure}[]
     \centering
     \begin{subfigure}[b]{0.3\textwidth}
              \centering
         \includegraphics[width=\textwidth]{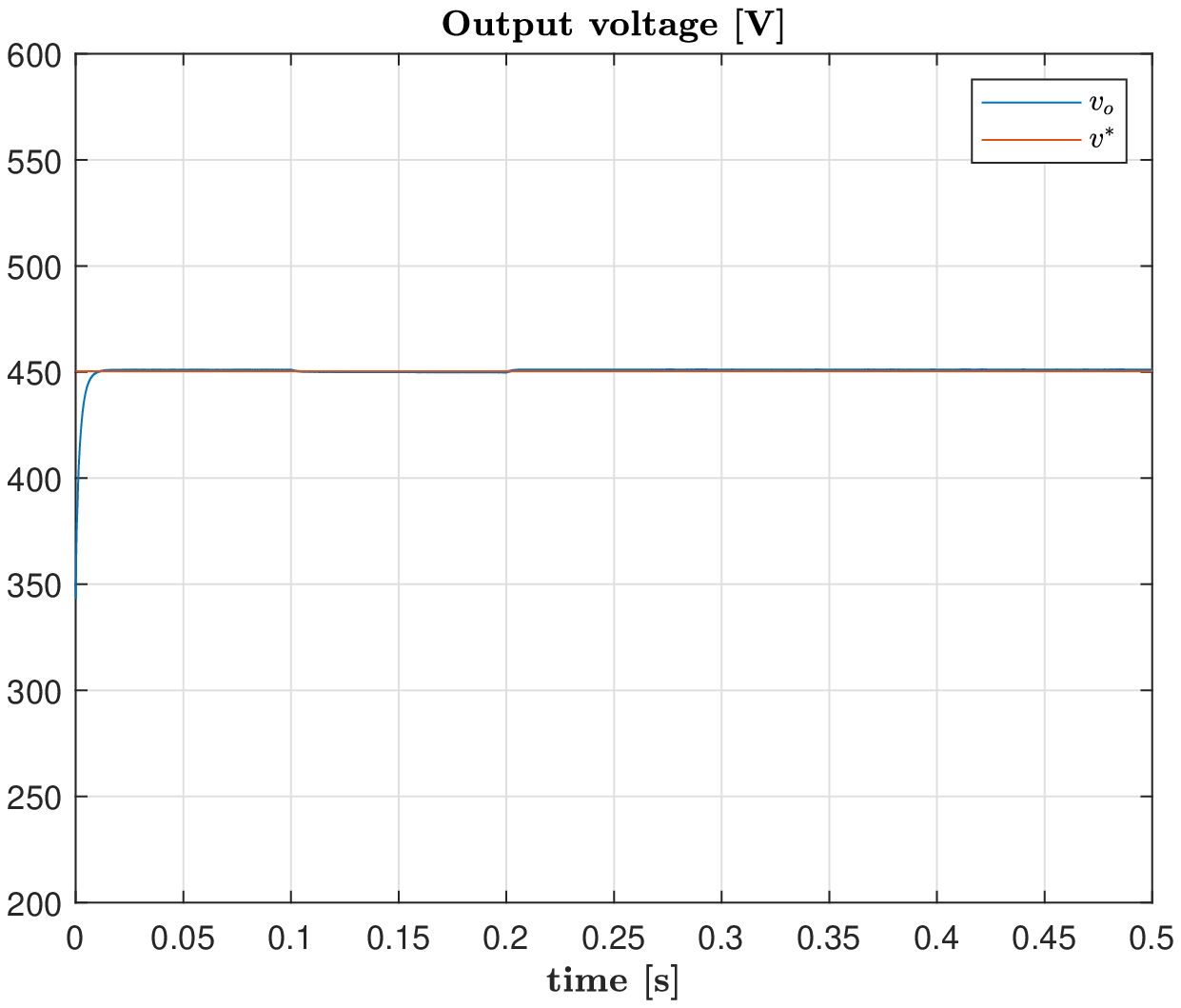}
         \caption{Output voltage}
         \label{fig:S3_output_voltage}
     \end{subfigure}\hspace{15mm}
       \begin{subfigure}[b]{0.3\textwidth}
     \centering
         \includegraphics[width=\textwidth]{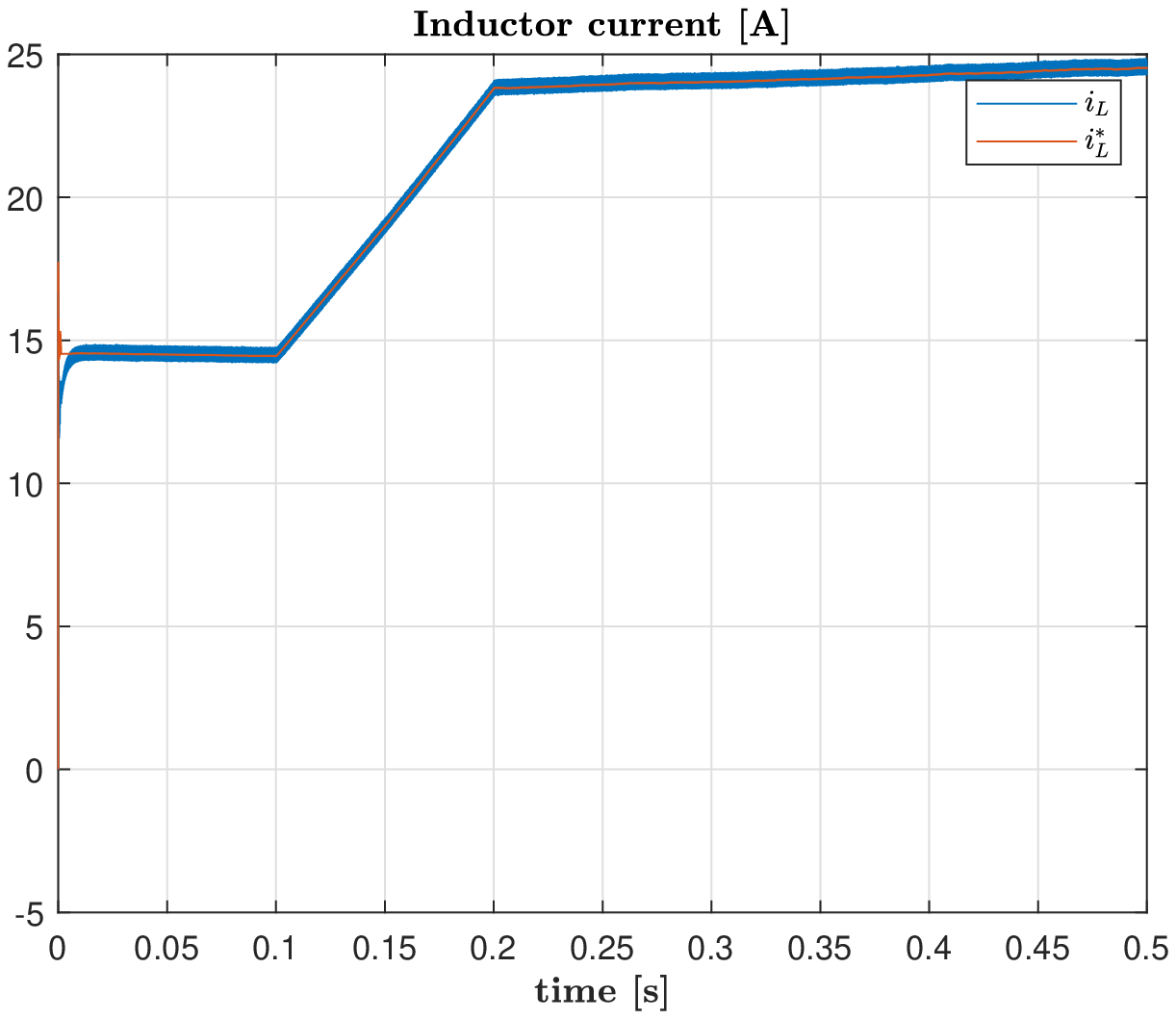}
         \caption{ Inductor current}
         \label{fig:S3_inductor_current}
     \end{subfigure}
          \\
          \begin{subfigure}[b]{0.3\textwidth}
         \centering
         \includegraphics[width=\textwidth]{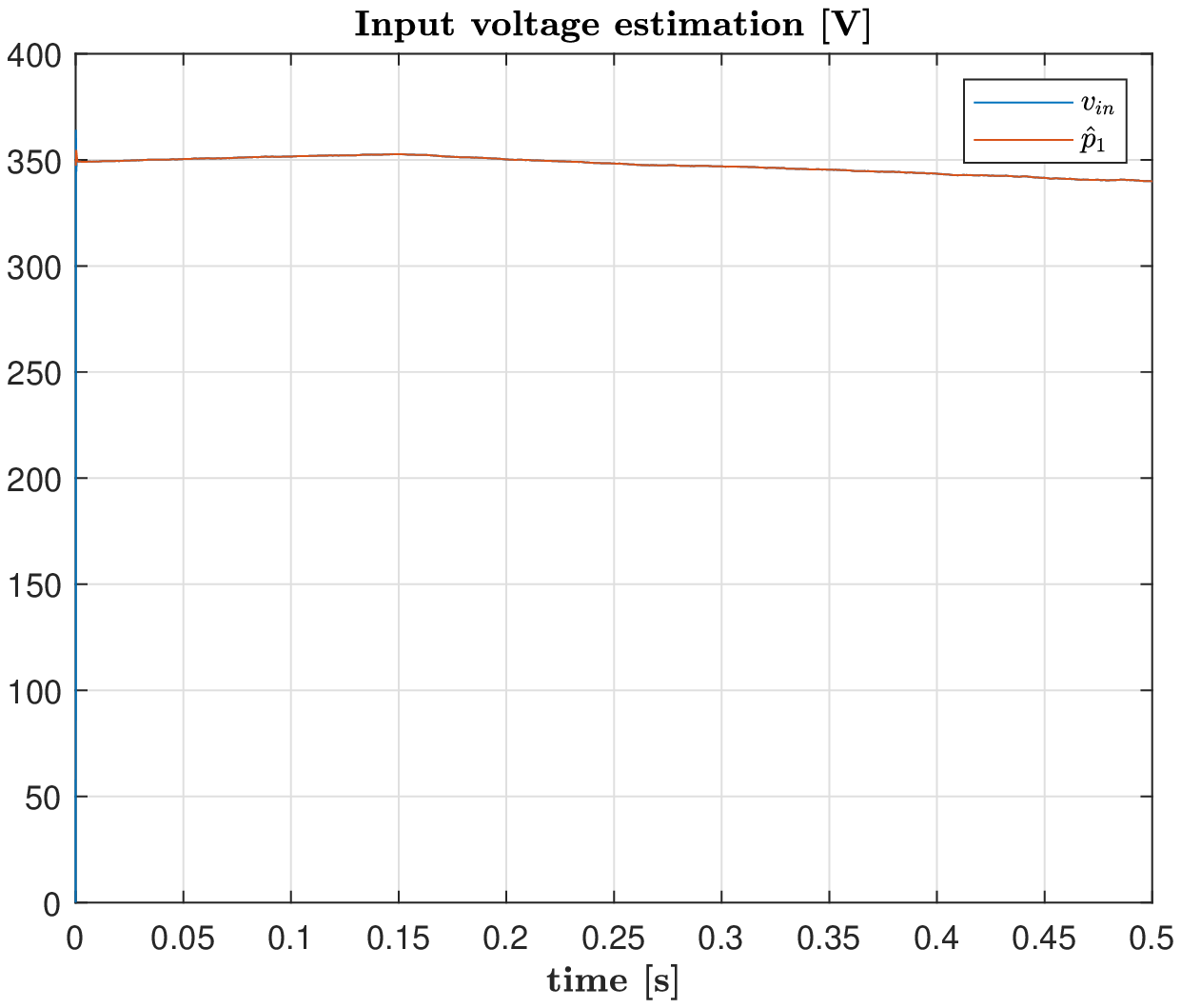}
         \caption{$v_{in}$ and $\hat{p}_1$}
         \label{fig:S3_p1_est}
     \end{subfigure}\hspace{15mm}
               \begin{subfigure}[b]{0.3\textwidth}
         \centering
         \includegraphics[width=\textwidth]{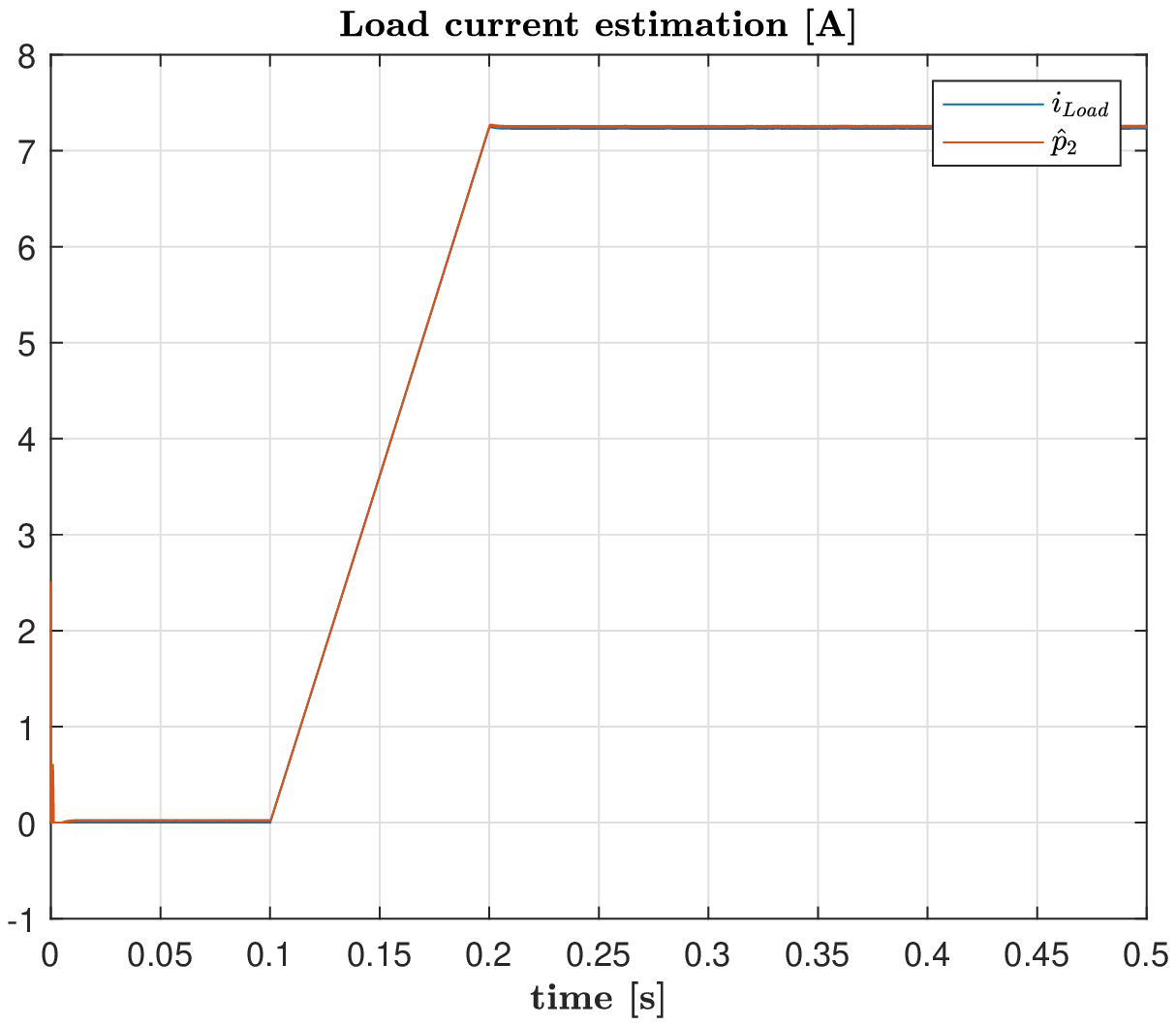}
         \caption{$i_{Load}$ and $\hat{p}_2$}
         \label{fig:S3_p2_est}
     \end{subfigure}
       \\
          \begin{subfigure}[b]{0.3\textwidth}
         \centering
         \includegraphics[width=\textwidth]{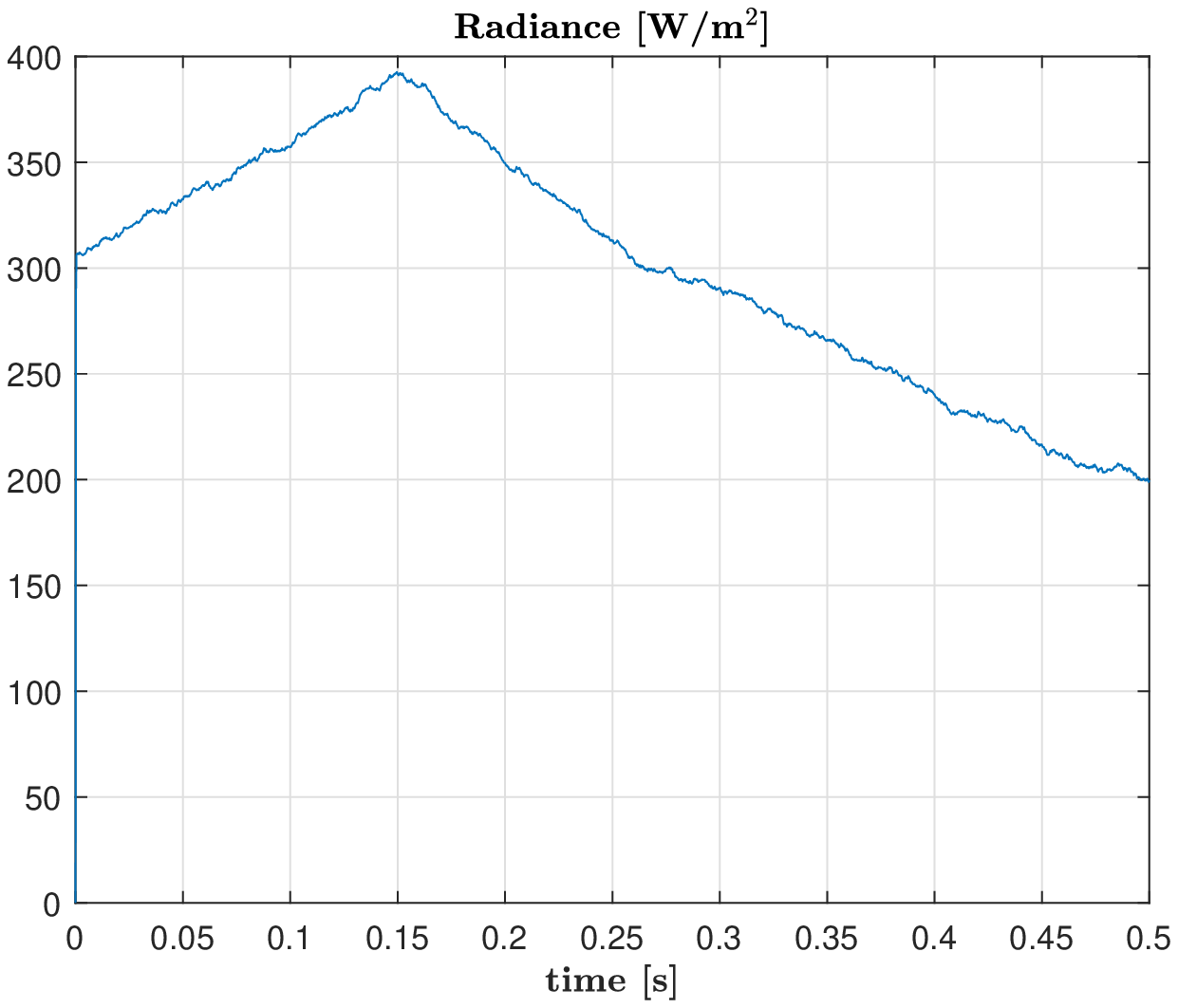}
         \caption{PV irradiance}
         \label{fig:S3_pv_irr}
     \end{subfigure}\hspace{15mm}
               \begin{subfigure}[b]{0.3\textwidth}
         \centering
         \includegraphics[width=\textwidth]{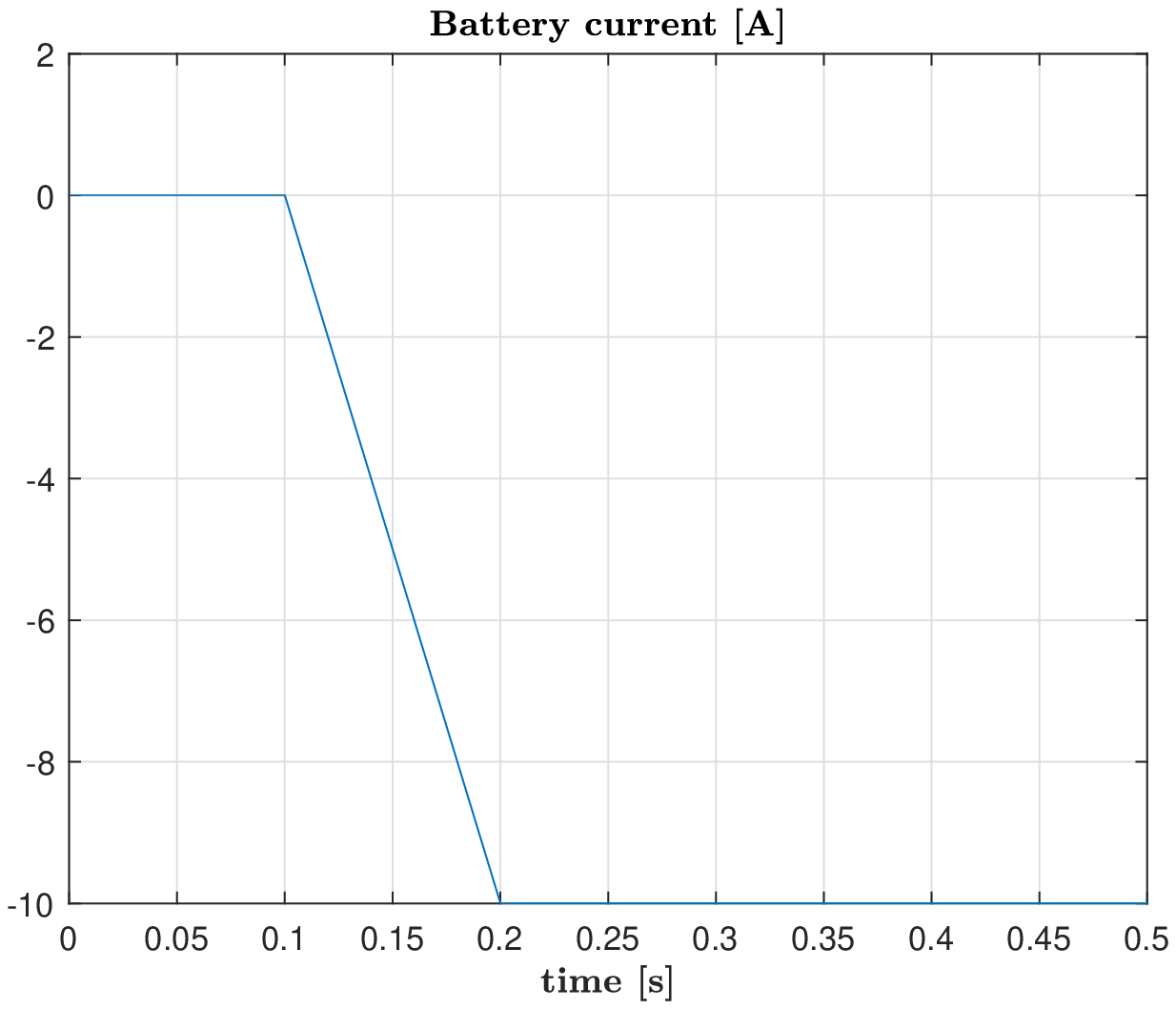}
         \caption{Battery current}
         \label{fig:S3_bat_current}
     \end{subfigure}
      \\
          \begin{subfigure}[b]{0.3\textwidth}
         \centering
         \includegraphics[width=\textwidth]{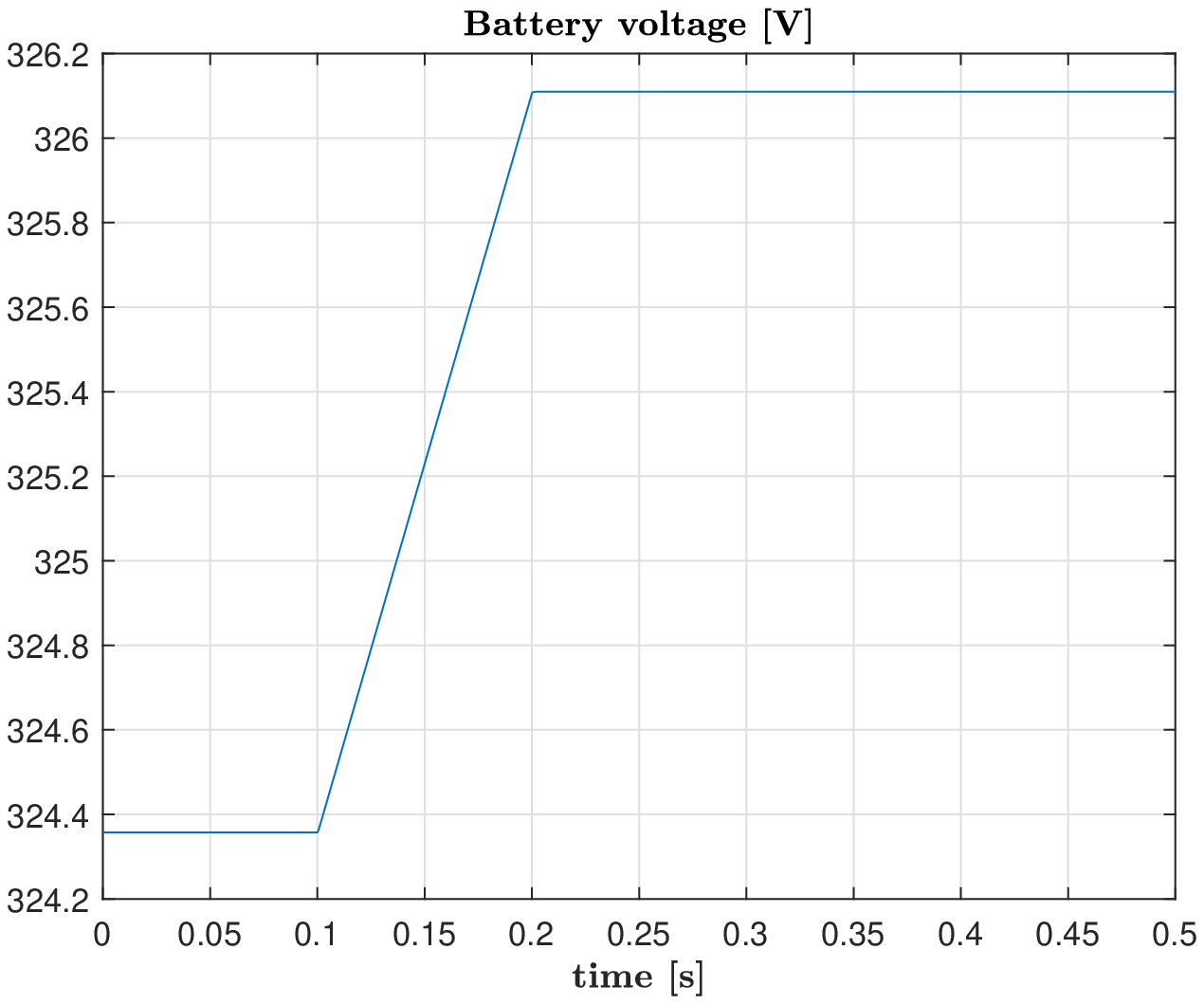}
         \caption{Battery voltage}
         \label{fig:S3_bat_volt}
     \end{subfigure}
     \hfill
        \caption{Simulation plots for \textbf{\textit{S3}}.}
        \label{fig:S3}
\end{figure}
 \subsubsection*{\textbf{[S1] Without parameter estimator $(\bm{\dot{p}}(t)=\textbf{0})$}} To illustrate the limitations of the switching controller in \eqref{eq:switchingLaw1} without equilibrium point update, a series of step changes in $\bm{p}(t)$ is applied in the following manner: $v_{in}=350V\xrightarrow{0.1s} 300V\xrightarrow{0.2s} 400V \xrightarrow{0.3s} 350V $ and $i_{Load}=0A\xrightarrow{0.15s} 20A\xrightarrow{0.25s} 10A \xrightarrow{0.3s} 0A$, where $i_{Load}$ is applied via a controlled current source block connected in parallel to the nominal resistance $R_o$. Fig. \ref{fig:S1} shows the obtained plots of the inductor current and output voltage over time. Note that, as soon as the first step disturbance is applied at $t=0.1$s, the system stabilizes at an equilibrium point which is different from the desired one. Moreover, at instant $t=0.15$s, the load disturbance is such that the new voltage equilibrium value is lower than the input (300V). Since this cannot be achieved in a DBC, the system stops switching and the voltage converges to 300V. The same thing happens at $t=0.2$s and $t=0.25$s. Then, in $t=0.3$s, with the step disturbances applied on the parameters $\bm{p}$, the nominal conditions are recovered and the switching controller is able to stabilize the system at the desired reference. This behavior is expected since the switching law ensures global stabilization, as noted in Section \ref{subsec:switchingLaw}. 
 \subsubsection*{\textbf{[S2] Idea Scenario $(\bm{\dot{p}}(t)=\textbf{0})$ with parameter estimator}}  Parametric perturbations in \textbf{\textit{S1}} are reapplied on the closed-loop system, but now with the estimator in \eqref{eq:estimator_noise} providing the required information to update $\bm{x^*}$ at each time-step. Simulation plots obtained in Fig. \ref{fig:S2} clearly indicate that $v_o(t)$ tracks the desired reference of $v_o^*=450V$ without any steady-state error as the designed estimator is able to accurately reconstruct the uncertain parameters using the state measurements and the mode $(\sigma)$ information which is subsequently used to update the equilibrium point $i_L^*$. Furthermore, the closed-loop system is able to recover from a simultaneous input voltage and load perturbations applied at $t=0.3s$. Also, since the assumption $\bm{\dot{p}}=\textbf{0}$ is satisfied, the estimates $\bm{\hat{p}}$ track the actual parameters asymptotically resulting in zero static error, which is evident from Fig. \ref{fig:S2_p1_est} and \ref{fig:S2_p2_est}. It is also worth noting that the estimates $\bm{\hat{p}}$ are not affected much by the measurement noise due to the additional filtering introduced in the designed estimator. 
 
\subsubsection*{\textbf{[S3] Practical scenario $(\bm{\dot{p}}(t)\neq\textbf{0})$ with parameter estimator for EV charging station with PV input}} In this scenario, a PV array is connected at the input side of the DBC with an input capacitor $(C_{in}=10\mu F)$ connected in parallel across the PV terminals and actual irradiation data (captured on-site at LAPLACE) with contracted time-scale is supplied to the PV array block.  On the load side, a Li-ion battery (50Ah, 300V) is connected via another DC-DC converter in order to ensure that the desired charging condition demanded by the EV user is satisfied. The charging starts at $t=0.1s$ with ramp type current having slope 0.1A/ms, which increases up to 10A till $t=0.2s$ as shown in Fig. \ref{fig:S3_bat_current} where charging current is considered as negative. Furthermore, the charging mode for the battery switches to constant voltage from constant current at $t=0.3s$, after which the charging current starts decreasing until it becomes zero close to full charge, however, the changes in current and voltage in different charging modes are too subtle to be notices on a short time-scale shown in Fig. \ref{fig:S3}. Note that the voltage  across the battery terminals and charging current is different from the output voltage across the converter terminals and reflected load current $i_{Load}$ as the battery is interfaced via another controlled DC-DC converter which ensures that the appropriate charging conditions are met. The PV irradiance, input voltage,  charging current for the battery and voltage across its terminals are all shown in Fig. \ref{fig:S3}. Simulation results obtained in Fig. \ref{fig:S3_output_voltage} to \ref{fig:S3_p2_est} illustrate the effectiveness of the designed control algorithm in tracking the desired reference despite the continuous fluctuations in the input and load parameters $(\bm{\dot{p}}(t)\neq\textbf{0})$.
\section{Conclusion}
\label{sec:conclusion}
A robust switching controller was proposed in this work for a DC-DC boost converter operating under input voltage and load perturbations. To address the problem of uncertain equilibrium point in the context of switching controllers, a parameter estimator was constructed under the assumption that the uncertain parameters are generated via a known linear exo-system. A sketch of the stability proof under ideal and practical conditions was provided for the estimator. Furthermore, the noise amplification and infinite switching frequency problems were addressed to facilitate practical implementation on real systems. Simulation results obtained for the ideal and practical test scenarios illustrate the effectiveness of the developed scheme. The designed estimator allows the flexibility to consider time varying parameters characterized by linear dynamics which covers a wider class of perturbations. For future works, it will be interesting to explore the effectiveness of designed estimator for different types of parametric variations and how higher-order time-polynomial approximations for such variations affect the estimation and closed-loop performance in the context of switching controller. Furthermore, closed-loop performance of the designed estimator with other types of switching controller, such as those designed for local stabilization,  can also be investigated.

\bibliographystyle{IEEEtran}
\bibliography{biblio}   

\begin{thebibliography}{10}
\providecommand{\url}[1]{#1}
\csname url@samestyle\endcsname
\providecommand{\newblock}{\relax}
\providecommand{\bibinfo}[2]{#2}
\providecommand{\BIBentrySTDinterwordspacing}{\spaceskip=0pt\relax}
\providecommand{\BIBentryALTinterwordstretchfactor}{4}
\providecommand{\BIBentryALTinterwordspacing}{\spaceskip=\fontdimen2\font plus
\BIBentryALTinterwordstretchfactor\fontdimen3\font minus
  \fontdimen4\font\relax}
\providecommand{\BIBforeignlanguage}[2]{{%
\expandafter\ifx\csname l@#1\endcsname\relax
\typeout{** WARNING: IEEEtran.bst: No hyphenation pattern has been}%
\typeout{** loaded for the language `#1'. Using the pattern for}%
\typeout{** the default language instead.}%
\else
\language=\csname l@#1\endcsname
\fi
#2}}
\providecommand{\BIBdecl}{\relax}
\BIBdecl

\bibitem{kumar2017DCmicrogrid}
D.~Kumar, F.~Zare, and A.~Ghosh, ``Dc microgrid technology: System
  architectures, ac grid interfaces, grounding schemes, power quality,
  communication networks, applications, and standardizations aspects,''
  \emph{IEEE Access}, vol.~5, pp. 12\,230--12\,256, 2017.

\bibitem{justo2013ac}
J.~J. Justo, F.~Mwasilu, J.~Lee, and J.-W. Jung, ``{AC-microgrids versus
  DC-microgrids with distributed energy resources: A review},'' \emph{Renewable
  and sustainable energy reviews}, vol.~24, pp. 387--405, 2013.

\bibitem{kaur2017state}
S.~Kaur, T.~Kaur, R.~Khanna, and P.~Singh, ``{A state of the art of DC
  microgrids for electric vehicle charging},'' in \emph{{2017 4th Int. Conf.
  Signal Proc., Comp. Control (ISPCC)}}.\hskip 1em plus 0.5em minus 0.4em\relax
  IEEE, 2017, pp. 381--386.

\bibitem{dahale2017overview}
S.~Dahale, A.~Das, N.~M. Pindoriya, and S.~Rajendran, ``{An overview of DC-DC
  converter topologies and controls in DC microgrid},'' in \emph{2017 7th Int.
  Conf. Power Systems (ICPS)}.\hskip 1em plus 0.5em minus 0.4em\relax IEEE,
  2017, pp. 410--415.

\bibitem{kobaku2017experimental}
T.~Kobaku, S.~C. Patwardhan, and V.~Agarwal, ``Experimental evaluation of
  internal model control scheme on a dc--dc boost converter exhibiting
  nonminimum phase behavior,'' \emph{IEEE Trans. Power Electron.}, vol.~32,
  no.~11, pp. 8880--8891, 2017.

\bibitem{kobaku2020quantitative}
T.~Kobaku, R.~Jeyasenthil, S.~Sahoo, R.~Ramchand, and T.~Dragicevic,
  ``Quantitative feedback design-based robust pid control of voltage mode
  controlled dc-dc boost converter,'' \emph{IEEE Trans. Circ. Sys. II: Express
  Briefs}, vol.~68, no.~1, pp. 286--290, 2020.

\bibitem{xu2020review}
Q.~Xu, N.~Vafamand, L.~Chen, T.~Dragi{\v{c}}evi{\'c}, L.~Xie, and F.~Blaabjerg,
  ``{Review on advanced control technologies for bidirectional DC/DC converters
  in DC microgrids},'' \emph{{IEEE J. Emer. Selec. Topics Power Electron.}},
  vol.~9, no.~2, pp. 1205--1221, 2020.

\bibitem{de2022switching}
R.~P. De~Souza, Z.~Kader, and S.~Caux, ``Switching control applied to
  interconnected boost converters: A comparison with hysteresis current
  control,'' in \emph{2022 International Symp. Power Electronics, Electrical
  Drives, Automation and Motion (SPEEDAM)}.\hskip 1em plus 0.5em minus
  0.4em\relax IEEE, 2022, pp. 541--546.

\bibitem{bayati2020sinusoidal}
M.~Bayati, M.~Abedi, G.~B. Gharehpetian, and M.~Farahmandrad,
  ``Sinusoidal-ripple current control in battery charger of electric
  vehicles,'' \emph{IEEE Trans. Vehic. Techn.}, vol.~69, no.~7, pp. 7201--7210,
  2020.

\bibitem{beneux2019adaptive}
G.~Beneux, P.~Riedinger, J.~Daafouz, and L.~Grimaud, ``Adaptive stabilization
  of switched affine systems with unknown equilibrium points: Application to
  power converters,'' \emph{Automatica}, vol.~99, pp. 82--91, 2019.

\bibitem{ndoye2022switching}
A.~Ndoye, R.~Delpoux, J.-F. Tr{\'e}gou{\"e}t, and X.~Lin-Shi, ``Switching
  control design for lti system with uncertain equilibrium: Application to
  parallel interconnection of dc/dc converters,'' \emph{Automatica}, vol. 145,
  p. 110522, 2022.

\bibitem{astolfi2021use}
D.~Astolfi, L.~Zaccarian, and M.~Jungers, ``On the use of low-pass filters in
  high-gain observers,'' \emph{Systems \& Control Letters}, vol. 148, p.
  104856, 2021.

\bibitem{Bolzern:2004}
P.~Bolzern and W.~Spinelli, ``Quadratic stabilization of a switched affine
  system about a nonequilibrium point,'' in \emph{Proc. 2004 American Control
  Conf.}, Boston, MA, USA, 2004, pp. 3890--3895.

\bibitem{chen2004disturbance}
W.-H. Chen, ``Disturbance observer based control for nonlinear systems,''
  \emph{IEEE/ASME Trans. Mech.}, vol.~9, no.~4, pp. 706--710, 2004.

\bibitem{ahmad2021active}
S.~Ahmad and A.~Ali, ``On active disturbance rejection control in presence of
  measurement noise,'' \emph{IEEE Trans. Ind. Electron.}, vol.~69, no.~11, pp.
  11\,600--11\,610, 2021.

\end{thebibliography}

\end{document}